\documentclass[11pt]{article}
\usepackage[margin=1.5in]{geometry}
\usepackage{amsmath}
\usepackage{mathrsfs}
\usepackage{amsfonts}
\usepackage{bbm}
\usepackage{graphicx}
\usepackage{moreverb}
\usepackage{mathtools}
\usepackage[font=small]{caption}
\usepackage{authblk}
\usepackage{url}

\usepackage{natbib}
\renewcommand{\bibsection}{} 

\newcommand{\argmax}{\operatornamewithlimits{argmax}}

\author[1]{Surojit Biswas\thanks{Correspondence: surojitbiswas@g.harvard.edu}}
\affil[1]{Department of Biomedical Informatics. Harvard Medical School. Boston, MA 02215. USA.}

\date{}

\title{The latent logarithm}

\begin{document}
\maketitle
\vspace{-10mm}
\begin{abstract}
Count or non-negative data are often log transformed to improve heteroscedasticity and scaling. To avoid undefined values where the data are zeros, a small pseudocount (e.g. 1) is added across the dataset prior to applying the transformation. This pseudocount considers neither the measured object's \emph{a priori} abundance nor the confidence with which the measurement was made, making this practice convenient but statistically unfounded. I introduce here the latent logarithm, or lag. lag assumes that each observed measurement is a noisy realization of an unmeasured latent abundance. By taking the logarithm of this learned latent abundance, which reflects both sampling confidence/depth and the object's \emph{a priori} abundance, lag provides a probabilistically coherent, stable, and intuitive alternative to the questionable, but conventional ``$\log(x + \textrm{pseudocount})$.'' \\
\\
\textbf{Availibility}: MATLAB code to compute the latent logarithm can be found here: \url{https://github.com/surgebiswas/latent_log.git}. 
\end{abstract}

\section{Introduction}
When working with count, or more generally, non-negative data one often encounters zeros. Furthermore, such data usually demonstrate heteroscedasticity, with variance increasing in mean -- a consequence of their governing probabilistic processes (e.g. Poisson, or Negative-Binomial sampling) \cite{Hilbe2011}. 

In order to render the data more homoscedastic, it is popular to log-transform the data prior to performing data analysis. However, as $\log(0)$ is undefined, applying this transformation directly across a dataset is not possible. One often therefore adds a small pseudocount (e.g. $+1$ when working with count data) prior to applying the log, but this is a questionable practice. Samples are often unevenly explored, and it can be unclear whether a zero actually suggests zero abundance, or some minuscule abundance that was not within the resolution of sampling. 

Indeed, the authors of \cite{OHara2010} argue against log-transforming count data prior to applying more standard, easy-to-work with statistical approaches (e.g. linear regression). Instead they argue for modeling the data directly using a Poisson or Negative Binomial GLM, both of which model the logarithm of the expected value of the data\footnote[2]{log transforming the data and applying linear regression is modeling the expected value of the log of the data, which is not the same as modeling the log of the expected value}.  I agree with these intuitions. Nevertheless, by virtue of modeling (log) expectations as a linear combination of user defined predictors, GLMs impose structure on the data and require it to be treated in a supervised manner. Thus, GLMs offer us little in the way of applying unsupervised methods (e.g. PCA, MDS, clustering) to non-negative data.

Using a Poisson-Normal hierarchical model, I propose here the \emph{latent logarithm}, hereafter ``lag'', that computes the logarithm of the measured object's latent, or denoised abundance in an unsupervised manner. Importantly, in the limit of data lag $=$ log, and in the absence of it, lag returns a prior belief. Furthermore, lag considers the level of confidence in or exploration of a sample, such that a zero for a sample that was well explored is treated differently than a zero for a sample for which we have low confidence. Thus the latent logarithm provides an intuitive and more nuanced alternative to the standard psuedocounted logarithm for application to count or non-negative data.

\section{Methods}

\subsection{Model}
Let $t \in \mathbb{R}_{\geq 0}^n$ be a $n$-samples long vector of data (e.g. counts, rates). Let $o \in \mathbb{R}_{> 0}^n$ be a $n$-samples long vector of `offsets', `exposures', sampling depths, or `confidences' as makes conceptual sense. For example, in the case where $t_i$ represents the number of times a specific species of animal was observed in a random sampling, $o_i$ would be the size of the random sample. 

I assume that associated with each $t_i$ there is a true, but unmeasured log-\emph{latent rate} $z_i$ that $t_i$ is ultimately a noisy realization of. Specifically, I assume the following hierarchical model,
\begin{align*}
z_i &\sim \mathcal{N}\left(\mu, \sigma^2\right) \\
t_i & \sim \textrm{Continuous-Poisson}(\exp(z_i)o_i)
\end{align*}
Here $\mu$ and $\sigma^2$ denote the mean and variance of $z_i$. As $z_i$ is the log-latent rate, $\exp(z_i)$ is therefore the \emph{latent rate} and $\exp(z_i)o_i$ is the \emph{latent abundance}. I define the Continuous-Poisson distribution to have the following density function with support on $x \in [0, \infty)$: 
\[
f(x|\lambda) = C_{\lambda}\frac{ \lambda^x e^{-\lambda}}{\Gamma(x + 1)}
\]
where $C_\lambda$ is a normalization constant that ensures the density integrates to unity. This distribution has the same shape and moments as the Poisson, but has the added advantage of being able to consider all non-negative data -- not just counts.

As an example, suppose we are looking for different species of birds. In this case, $t_i$ would represent the number of times we saw the bird on excursion $i$, and $o_i$ would represent the number of hours we spent looking. For a rare bird, the latent rate $\exp(z_i)$ would be close to zero sightings/hour (negative $z_i$), whereas for a common bird, $\exp(z_i)$ would be a positive number (positive $z_i$). From this example it's clear that we have,
\[
\textrm{units}(\textrm{latent rate}) = \textrm{units}(\exp(z_i)) = \frac{ \textrm{units}(t_i)}{ \textrm{units}(o_i)}
\]

With these intuitions in mind, I define the \emph{latent logarithm} to be,
\[
\textrm{lag}(t_i) = \log(\mathbb{E}[t_i|z_i]) = z_i + \log(o_i)
\]
Given we are often more directly interested in the rate of an event (e.g. seeing a bird many times when looking for many hours is the same as seeing a bird only a few times when looking for only a few hours), I also define the \emph{normalized latent logarithm} (nlag) to be, 
\[
\textrm{nlag}(t_i) = \log(\mathbb{E}[t_i|z_i]/o_i) = z_i
\]
In practice, we will not know $z_i$, $\mu$, and $\sigma^2$ and must therefore learn their value.

\subsection{Inference}

The joint likelihood of $\{t_i\}_{i=1}^n$ and $\{z_i\}_{i=1}^n$ is given by,
\begin{align*}
p(\{t_i\}_{i=1}^n, \{z_i\}_{i=1}^n | \mu, \sigma^2) & = p(\{t_i\}_{i=1}^n | \{z_i\}_{i=1}^n, \mu, \sigma^2)  p(\{z_i\}_{i=1}^n |\mu, \sigma^2)\\
& = \prod_{i=1}^n p(t_i | z_i, \mu, \sigma^2) p(z_i | \mu, \sigma^2) \\
& = \prod_{i=1}^n C_{\exp(z_i)o_i} \frac{ [\exp(z_i)o_i]^{t_i} e^{-\exp(z_i)o_i}}{\Gamma(t_i + 1)} \\ & \qquad \times \frac{1}{\sqrt{2\pi\sigma^2}} \exp\left( -\frac{1}{2\sigma^2}(z_i - \mu)^2 \right) 
\end{align*}
This objective is difficult to optimize jointly in $z$ and in $\mu$ and $\sigma^2$. However, given $z$ the task is considerably easier. Our estimates for $\mu$ and $\sigma^2$, would simply the usual sample mean and variance of $z$. This suggests an approach where we iteratively condition on some estimate or distribution over $z$ given estimates $\hat{\mu}$ and $\hat{\sigma}^2$, and subsequently maximize $\mu$ and $\sigma^2$ given our current understanding of $z$.

One approach could be to use Expectation-Maximization (EM), in which we continuously maximize the expected log-likelihood. Because this expectation is taken with respect to $z$, we require a posterior distribution over $z$, which in turn requires us to calculate the marginal distribution of $p(t_i | \mu, \sigma^2) = \int p(t_i | z_i, \mu, \sigma^2)p(z_i | \mu, \sigma^2) \textrm{d}z_i$. Unfortunately, this integral is not analytically solvable, making an exact EM approach hard.

While we could resort to inexact sampling techniques (e.g. using Metropolis-Hastings MCMC, where the marginal distribution is not needed to sample from the posterior), these approaches are slow. Therefore, instead, I consider $z$ to simply be another variable in the model, and employ the iterative conditional modes (ICM) algorithm \cite{Besag1986}. As mentioned before, the estimates of $\mu$ and $\sigma^2$ are straightforward given an estimate of $z$. Given $\mu$ and $\sigma^2$, our goal will then be to set $z$ to be the maximizer of its posterior distribution. These two steps can then be iterated until convergence of the (log) data likelihood given above.

\subsubsection{Maximum \emph{a posteriori} estimation of $z$ given $\mu$ and $\sigma^2$}

A reasonable estimate of $z_i$ is given by the mode value of its posterior distribution. This objective is given by,
\begin{align*}
\hat{z}_i = \argmax_{z_i} p\left(z_i|t_i,\mu, \sigma^2 \right) & = \argmax_{z_i} \log \left[ \frac{ p\left(t_i|z_i,\mu, \sigma^2 \right)p\left(z_i |\mu, \sigma^2 \right) }{ \int_{-\infty}^{\infty} p\left(t_i|x,\mu, \sigma^2 \right)p\left(x |\mu, \sigma^2 \right) \textrm{d}x } \right] \\
& = \argmax_{z_i} \log  p\left(t_i|z_i,\mu, \sigma^2 \right) + \log p\left(z_i |\mu, \sigma^2 \right)  \\
& = \argmax_{z_i} \log\left\{ C_{\exp(z_i)o_i} \frac{ [\exp(z_i)o_i]^{t_i} e^{-\exp(z_i)o_i}}{\Gamma(t_i + 1)}\right\} \\ 
& \hspace{13.5mm} + \log\left\{ \frac{1}{\sqrt{2\pi\sigma^2}} \exp\left( -\frac{1}{2\sigma^2}(z_i - \mu)^2 \right) \right\} \\
& = \argmax_{z_i} t_i z_i - \exp(z_i)o_i -\frac{1}{2\sigma^2}(z_i - \mu)^2 
\end{align*}
where I have progressively dropped constants that do not depend on $z_i$. 

Differentiating we get, 
\[
\frac{\partial}{\partial z_i} t_i z_i - \exp(z_i)o_i -\frac{1}{2\sigma^2}(z_i - \mu)^2 = t_i - \exp(z_i)o_i -\frac{z_i - \mu}{\sigma^2}.
\]
We cannot solve for this gradient analytically, so instead we rely on Newton-Raphson to optimize the gradient numerically. The required Hessian is given by,
\[
\frac{\partial}{\partial z_i} t_i - \exp(z_i)o_i -\frac{z_i - \mu}{\sigma^2} = - \exp(z_i)o_i -\frac{1}{\sigma^2} < 0 .
\]
Note that the Hessian is negative-definite, which implies there is a unique maximum for our objective.  

\subsection{Properties}

I now note some important properties of the latent logarithm. Note that given some moderate $z_i$, $t_i \rightarrow \infty$ as $o_i \rightarrow \infty$ since $\mathbb{E}[t_i|z_i] = \exp(z_i)o_i$. Thus in the limit of data we have, 
\begin{align*}
&  \lim_{t_i \to \infty, o_i \to \infty} \argmax_{z_i} t_i z_i - \exp(z_i)o_i -\frac{1}{2\sigma^2}(z_i - \mu)^2 = \argmax_{z_i} t_i z_i - \exp(z_i)o_i
\end{align*}
which is simply an optimization of the Poisson likelihood. With one data point the maximum likelihood estimator of a Poisson mean is just the value of the datum itself. Consequently, 
\[
\lim_{t_i \to \infty,o_i \to \infty}\textrm{lag}(t_i) = \log(\mathbb{E}(t_i|z_i)) = \log(t_i).   
\]

Now consider the situation in which again we have ample observation ($o_i \rightarrow \infty$), but that $t_i = 0$. Intuitively, this must be because the latent rate is minuscule. To confirm this intuition mathematically, notice that in this case our objective becomes, 
\begin{align*}
&  \lim_{t_i \to 0, o_i \to \infty} \argmax_{z_i} t_i z_i - \exp(z_i)o_i -\frac{1}{2\sigma^2}(z_i - \mu)^2 = \argmax_{z_i} - \exp(z_i)o_i
\end{align*}

Note that $- \exp(z_i)o_i$ has no maximum, but that it is always increasing as $z_i \rightarrow -\infty$. This confirms our intuitions, and we can see that when samples are deeply explored but still no events are detected, lag (or more appropriately in this case, nlag), will suggest the rate of the process is low.

Consider the case where we have limited data and observation (e.g. $t_i = 0$ and $o_i = 0$). Here we have, 
\[
\lim_{t_i \to 0, o_i \to 0} \argmax_{z_i} t_i z_i - \exp(z_i)o_i -\frac{1}{2\sigma^2}(z_i - \mu)^2 = \argmax_{z_i} -\frac{1}{2\sigma^2}(z_i - \mu)^2
\]
which is simply the Normal likelihood, for which $z_i = \mu$ is the maximizer. Thus,
\[
\lim_{t_i \to 0,o_i \to 0}\textrm{nlag}(t_i) = \log(\mathbb{E}(t_i|z_i)/o_i) = \mu.   
\]
This says that in the absence of data, our best guess for the (log) latent rate of an event, is simply the mean of the distribution that encodes prior beliefs about the process rate.

Finally, consider the influence of the prior. We can write the objective as follows:
\[
\argmax_{z_i} \underbrace{[t_i z_i - \exp(z_i)o_i]}_{\textrm{Poisson component}} + \underbrace{\left[-\frac{1}{2\sigma^2}(z_i - \mu)^2\right]}_{\textrm{Normal component}}
\]
As the prior variance decreases, thereby encoding stronger prior beliefs, the influence of the `Normal component' of the posterior increases. Similarly, as the prior mean shifts toward extreme values (again encoding stronger prior beliefs), the `Normal component' of the posterior again dominates and accordingly ``drags'' along with it the estimate of $z_i$.  

\section{Results}

I first examine the latent logarithm when $\mu$ and $\sigma$ are fixed in order to gain a better understanding of how the Normal prior distribution interacts with the Continuous-Poisson layer. I then demonstrate complete use of the latent logarithm, where $z$, $\mu$, and $\sigma^2$ are all learned. 

\begin{figure}[t!]
\centering
\includegraphics[trim={1cm 6cm 1cm 7cm},clip,width=\textwidth]{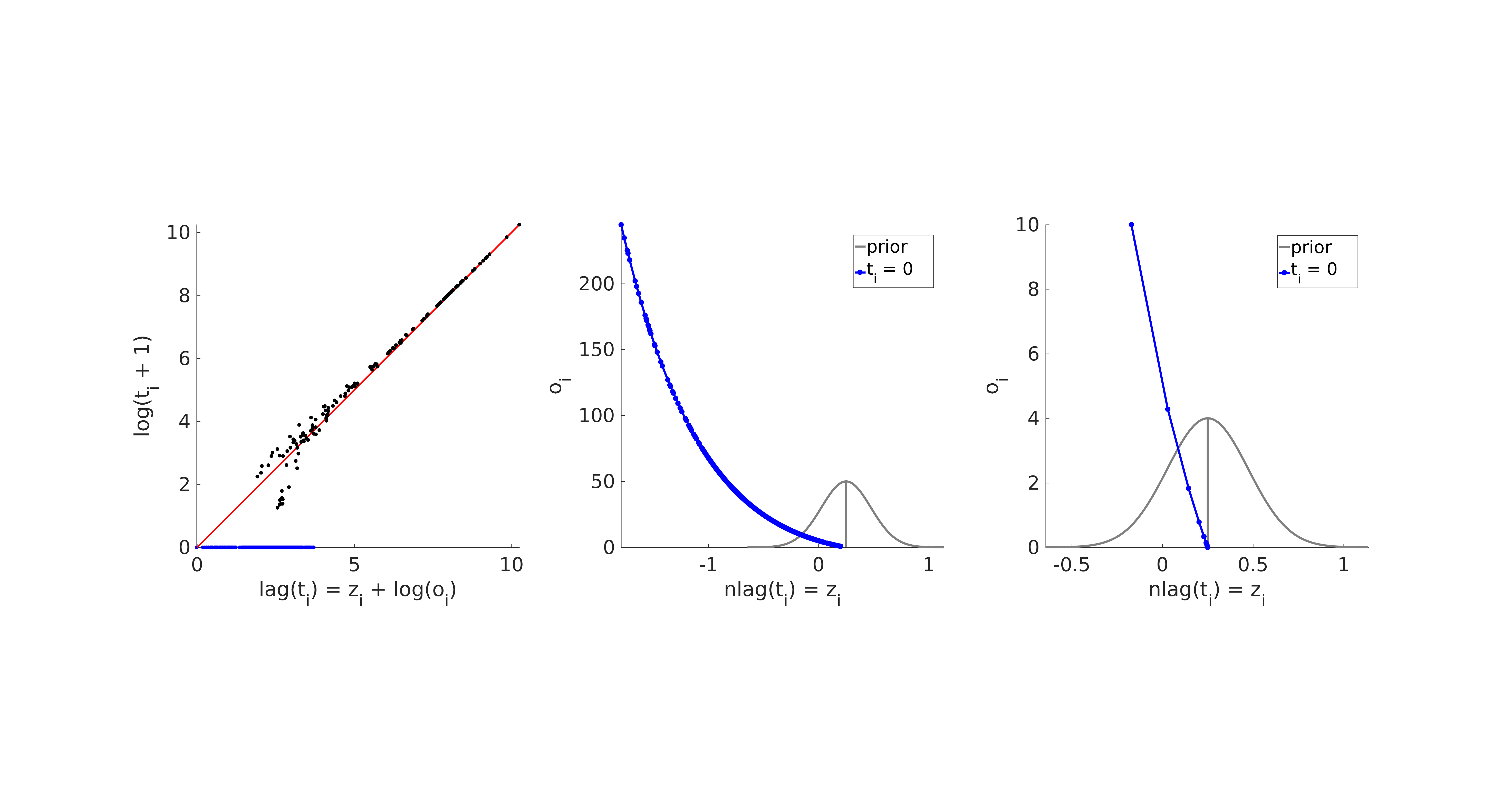}
\caption{Behavior of the latent logarithm on a gene expression dataset. left) Comparison of $\textrm{lag}(t_i)$ versus a conventional psuedocount approach of taking $\log(t_i + 1)$. The red line corresponds to the $x = y$ line. Points in blue are those for which $t_i$ equals 0. middle) Estimated log-latent TPM ($x$-axis) as the sequencing depth ($y$-axis) decreases for samples for which the gene was not detected. right) Estimates of log-latent TPMs for synthetic data where sequencing depth is shrunk to effectively zero. In the middle and right plots the gray curve illustrates the prior density.}
\end{figure}

\subsection{Fixed $\mu$ and $\sigma^2$}
In order to examine the behavior of the latent logarithm under a fixed prior, I investigated gene expression data consisting of 2597 samples for the \emph{Arabidopsis thaliana} gene AT2G43386 \cite{Biswas2016}. This gene is rarely expressed and in most of these samples has a measured abundance in Transcripts per Million reads (TPM) of 0. However, some of these 0 TPM samples were explored more heavily\footnote[3]{Gene expression is often measured by RNA sequencing, which for the purposes of this discussion, incompletely samples a large, heterogeneous pool of transcripts (copies of expressed genes). The number of detected transcripts for a specific gene is therefore proportional to the sequencing \emph{depth} of that sample -- or in other words, how deeply it was probed.} than others, and for the remaining samples a positive TPM was measured. I hereafter refer to this gene as the `rare' gene.

To obtain $t$, I first converted the measured TPMs to a measured number of transcripts by multiplying the measured TPM by the sequencing depth of the sample (in number of reads) divided by 1 million. Intuitively then, $o$ is simply the sequencing depths of each sample (in millions of reads). Note that given the original measurements were already given as rates (TPMs), we could directly model these rates in our model by setting $o_i = 1$ for all $i$. However, doing so would ignore the information encoded in the sequencing depth. For example, a TPM of 0 (indicating no expression) in a lightly sequenced sample is not as believable as a TPM of 0 in a heavily sequenced sample. To clearly demonstrate the influence of the prior distribution, I set the prior mean $\mu$ to $0.25$ and the the prior variance to $0.05$. 

Figure 1 illustrates the results of applying the latent logarithm to the above gene expression dataset. Specifically, Figure 1 (left) shows how $\textrm{lag}(t_i)$ compares to $\log(t_i + 1)$. The psuedocount of 1, a commonly added value, was added so that 0 values were not undefined. Notice that as $t_i$ grows large $\log(t_i + 1) \approx \log(t_i) \approx \textrm{lag}(t_i)$, as expected, and the prior has little influence. Note that the positive bias (above the red line) observed in the plot for moderate (non-zero), but not large $t_i$ is due to the effect of the pseudocount. 

Additionally, we see some variation between $\textrm{lag}(t_i)$ and $\log(t_i + 1)$ for when $t_i$ is moderate, but not large (e.g. around where $\log(t_i + 1) = 1.5$). For these samples lag estimates that the gene is actually more abundant than measured. Indeed, this is because these samples were not heavily sequenced, and so lag does not know whether the low measured abundance was due to variability introduced by shallow sequencing or because the gene is truly not very expressed. Consequently, it places some emphasis on the prior.

How lag manages the trade-off between trusting the data versus the prior is most clearly seen for samples in which the measured TPM abundance equals 0 (blue points in Figure 1 (left)). Intuitively, we should be most confident in such an estimate when our sequencing depth is high. This should be reflected in the fact that the estimated latent rate should be low. Conversely, if we measured an abundance of 0 but did not sequence very deeply, then we do not have much information to tell us whether that 0 really represents zero abundance or a zero due to technical error. Therefore, our best estimate for the estimated latent rate should simply reflect our prior belief. 

Indeed, this is exactly what we see in Figure 2 (middle), where I have plotted the estimated log-latent TPM ($\textrm{nlag}(t_i)$) along with the sequencing depth of that sample. Here, as sequencing depth increases, the latent logarithm is more and more confident that a measured abundance of 0 corresponds to a small log-latent TPM. Conversely, as depth decreases, the latent logarithm puts greater trust in the prior.

To further confirm the latent logarithm behaves as expected and is numerically stable for even the most scant data situations, I generated synthetic data points where $t_i = 0$ and $o_i$ ranged from $10$ to $1 \times 10^{-6}$ million reads on a logarithmic grid. The case where 0 transcripts are detected for a rare gene when a sample is sequenced to a depth of only 1 read is functionally equivalent to never probing that sample in the first place. Thus, the log-latent rate should converge to the prior mean. Figure 1 (right) confirms this intuition exactly.

\subsection{Complete usage}
\begin{figure}[t]
\centering
\includegraphics[trim={1cm 3cm 1cm 2cm},clip,width=\textwidth]{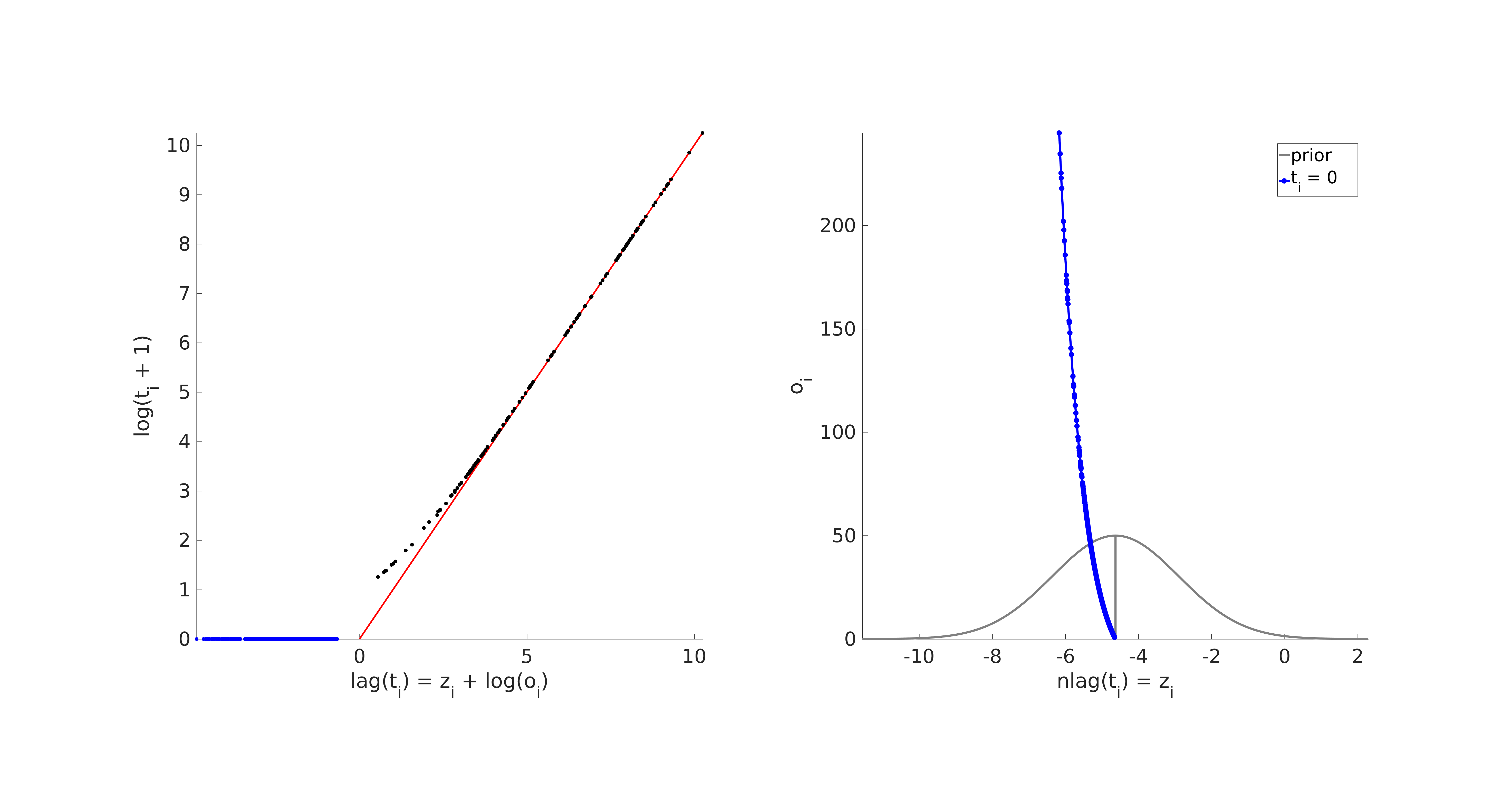}
\caption{Behavior of the latent logarithm for the `rare' gene. The left and right panels are presented in the same style as the left and middle panels of Figure 1.}
\end{figure}

In practice, we do not know $\mu$ and $\sigma^2$, and must learn it from the data. We initialize the prior mean and variance for the ICM algorithm using the $+1$ pseudocount approach such that $\hat{\mu}_{\textrm{init}}$ and $\hat{\sigma^2}_{\textrm{init}}$ equals the sample mean and variance of $\log(t/o + 1)$ (here, the division is taken element wise).

Figure 2 illustrates the completely estimated ($z$, $\mu$, $\sigma^2$ all learned) lag function for the `rare' gene. The learned latent mean and variance are $-4.63$ and $2.30$, which define a latent abundance distribution that is considerably left shifted and wider than the fixed-prior example given above (Figure 2, right). Stated another way, lag \emph{a priori} believes for this gene that an observed TPM of 0 is really a TPM more like $\exp(-4.63) = 0.0098$. If, as depth increases to a large value, the observed TPM is still 0, then lag estimates the latent TPM to be approximately $\exp(-6) = 0.0025$. Note that these values are a few orders of magnitude less than 1. So, whether one interprets the $+1$ pseudocount as a prior belief of the transcript's abundance or applies it out of convenience, it's clear that for this rarely expressed gene (TPM equals 0 94$\%$ of the time), a pseudocount of 1 is far too generous. This is made clear in examining Figure 2 (left) where $\log(t_i + 1)$ consistently exceeds $\textrm{lag}(t_i)$ for small $t_i$.  

I repeated this analysis for a more abundant gene, AT1G14630, which I hereafter simply refer to as the `abundant gene.' Unlike the rare gene, the abundant gene has non-zero TPM 97$\%$ of the time in the dataset. As expected, the learned latent abundance prior distribution is right shifted compared to the rare gene, with mean and variance equal to 0.83 and 1.80, respectively (Figure 3, right). Additionally, as seen before in previous examples, for samples with zero TPM levels the estimated latent abundance decreases in increasing sampling depth. However curiously, even for a shallowly sequenced sample (2.1096 million reads) in which we observe 0 TPM, the estimated log-latent TPM is still far from the prior mean. This is because for this abundant gene, 2 million reads should still be enough to detect it and so it's more likely that the observed 0 is real. lag hedges its bets by assigning this sample a latent TPM of 0.43.

Finally, note that unlike for the rare gene, $\textrm{lag}(t_i)$ generally (though not always when sufficient sampling is available) exceeds $\textrm{log}(t_i)$, especially for 0 TPM samples. Intuitively, this is simply because for a robustly expressed gene, with an average number of transcripts of 100.2 and a median transcript count of 49.8, a pseudocount of 1 is too stingy, especially for deeply sequenced samples.

\begin{figure}[t]
\centering
\includegraphics[trim={1cm 3cm 1cm 2cm},clip,width=\textwidth]{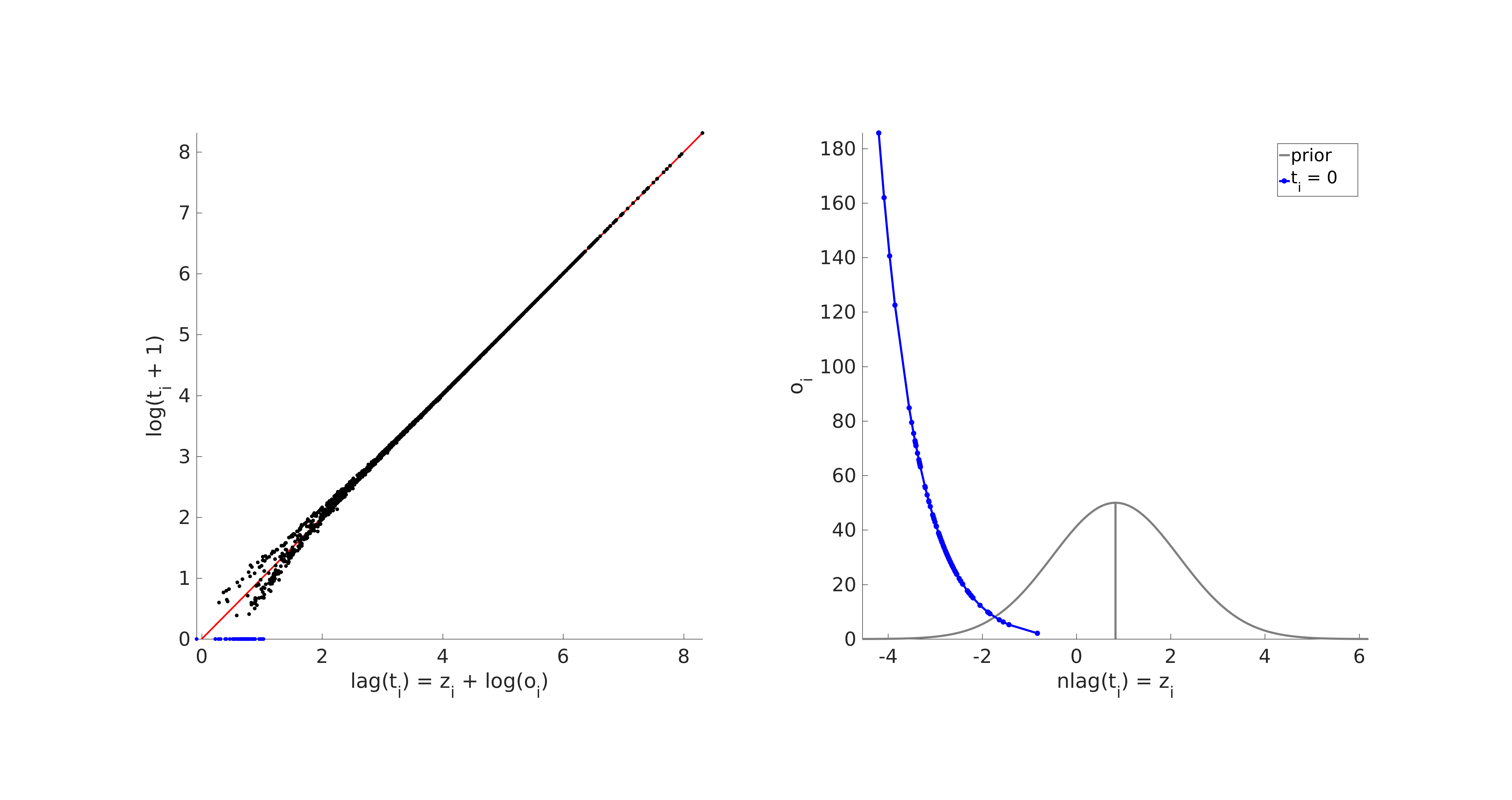}
\caption{Behavior of the latent logarithm for the `abundant gene.' The left and right panels are presented in the same style as the left and middle panels of Figure 1.}
\end{figure}

\section{Discussion}

In count or non-negative data, there are different kinds of zeros. Zero measured abundances are more believable for \emph{a priori} rare objects/events or deeply probed samples, than for abundant objects/events or shallowly probed samples. Applying a fixed pseudocount across the dataset ignores this information. lag addresses this shortcoming by taking the logarithm of the object's learned latent abundance, which considers both the object's overall abundance and sampling depth.

Nevertheless, unlike log, lag is a statistical routine, and like all methods of inference, it is only as good as the data it is applied to. Though our algorithm converges even for two samples, the estimated latent prior distribution and the latent abundances are not necessarily stable. In these situations, it may be prudent to constrain lag by placing prior distributions on the parameters of latent prior density. Alternatively and perhaps ideally, the latent prior density should be learned on a larger previously assembled dataset that's representative of the dataset to be analyzed. Importantly, when considering multiple intellectually linked datasets, the same latent prior distribution should be used so that lag values are globally comparable. This can be done by using an already learned prior (from a previous dataset), or by applying lag to a concatenation of all datasets under consideration.

Finally, one may question the use of the Normal distribution as a prior density over the log of the Poisson rate parameter\footnote[4]{or equivalently, the log-normal distribution over the Poisson rate}, instead of a more standard distribution, e.g. Gamma, which is conjugate to the Poisson. I argue the Normal actually enables greater expressivity. It's known that the Poisson-Normal hierarchy allows for the same mean-variance flexibility as the Negative-Binomial\footnote{Let $t_i \sim \textrm{Poisson}(\lambda)$ and $\lambda \sim \textrm{Gamma}(\alpha, \beta)$. Then $p(t_i|\alpha,\beta) = \int_\lambda p(t_i|\lambda)p(\lambda|\alpha,\beta)\textrm{d}\lambda$ is a Negative-Binomial distribution.} \cite{Aitchison1989}. However, unlike what can be done with the Gamma prior, one can encode considerable model structure into the mean and (co)variance of the Normal, as well as take advantage of its nice conditioning and marginalization properties.  For example, one might consider extending $\mu$ to be a conditional mean $\mu(X) = X\beta$, which depends on some user known/input set of covariates collected in the columns of $X$. Perhaps most excitingly, the univariate normal distribution may be extended to the multivariate domain, in which now the Normal's covariance matrix can couple latent abundances -- and therefore ultimately the observed abundances -- across multiple count measured objects \cite{Biswas2015, Aitchison1989}. Thus, the Normal-Poisson hierarchy can enable the application of very well established and powerful multivariate methods to count data.  

Though lag does not attempt this complexity, it provides a simple and sound transformation that can serve as an initialization or null model for these more expressive models. Importantly, it also offers a more principled way to preprocess non-negative data for unsupervised learning tasks than does the usual pseudocounted logarithm.

\section{Conclusion}

I introduced lag, a probabilistic alternative to the standard ``$\textrm{log}(x + \textrm{pseudocount})$'' that computes the logarithm of the measured object's learned latent abundance. In the limit of increasing data, lag quickly converges to log. However, in situations of poor sampling or object rarity,  lag considers all available information to calculate the object's log-latent abundance. This statistically sound and more nuanced approach is an improvement over the commonplace practice of transforming count-data-plus-pseudocount with the standard logarithm.

\section{References}
\bibliography{references}

\end{document}